\documentclass[a4paper,11pt]{article}
\usepackage{pos}
\usepackage{wrapfig}
\usepackage[export]{adjustbox}
\usepackage{enumitem}

\title{The EUSO-SPB2 Cherenkov Telescope - Flight Performance and Preliminary Results
}
 \ShortTitle{}

\author*[a]{Eliza Gazda}

\affiliation[a]{School of Physics \& Center for Relativistic Astrophysics, Georgia Institute of Technology,\\
		     837 State Street NW, Atlanta, GA 30332-0430, USA}

\onbehalf{for the JEM-EUSO collaboration}

\emailAdd{egazda6@gatech.edu}

\abstract{Astrophysical Very-High-Energy (VHE, >10 PeV) neutrinos deliver crucial information about the sources of Ultra-High-Energy Cosmic Rays (UHECRs), the composition of UHECRs, and neutrino/particle physics at highest energies.  UHE-tau neutrinos skimming the Earth’s surface produce tau leptons, which can emerge from the ground, decay, and start an upward-going extensive air shower (EAS) in the Earth’s atmosphere. The tau neutrino can be reconstructed by imaging the EAS. We developed an atmospheric Cherenkov Telescope flying on the Extreme Universe Space Observatory Super Pressure Balloon 2 (EUSO-SPB2) mission to test the air-shower imaging concept at high altitudes. The EUSO-SPB2 ultra-long-duration balloon mission is a precursor of the Probe of Extreme Multi-Messenger Astrophysics (POEMMA), a candidate for an astrophysics probe-class mission. The telescope implements Schmidt optics with a 0.785 $m^2$ light collection area and a 512-pixel SiPM camera covering a 12.8° x 6.4° (Horizontal x Vertical) field-of-view with 0.4° resolution. The camera signals are sampled with 100 MSa/s and digitized with 12-bit resolution. The objectives of the EUSO-SPB2 Cherenkov telescope include a search for UHE neutrinos below Earth’s limb, UHECRs above the limb, the study of the night sky background, and studying the telescope's performance. In this presentation, I will present an overview of the Cherenkov telescope and discuss the in-flight performance of the telescope.}

\ConferenceLogo{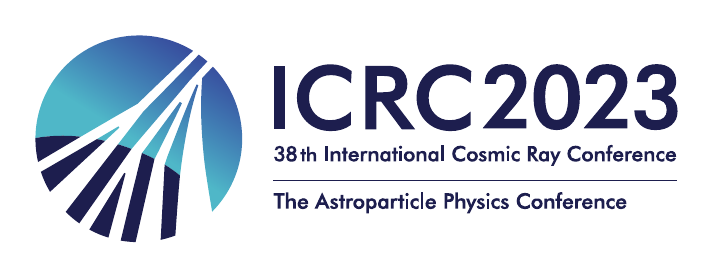}

\FullConference{%
The 38th International Cosmic Ray Conference (ICRC2023)\\
  26 July -- 3 August, 2023\\
  Nagoya, Japan}

\begin{document}
\maketitle

\section{Introduction}

The search for VHE and UHE neutrinos is part of Multi-Messenger Astronomy which is a rapidly advancing field due to recent groundbreaking discoveries such as the detection of the astrophysical neutrino flux by IceCube \cite{IceCube2018} and the observation of gravitational waves by LIGO \cite{LIGO2016}. Future instruments will further explore the Very-High-Energy (VHE) neutrino window (>10 PeV)  such as The Probe of Extreme Multi-Messenger Astrophysics (POEMMA) \cite{POEMMA-JCAP}, Terzina on board NUSES \cite{Terzina}, and Trinity \cite{Trinity}. All these instruments use the imaging atmospheric Cherenkov technique (IACT).
\begin{wrapfigure}{R}{0.4\textwidth}
\vspace{-.45cm}
    \centering
      \captionsetup{width=0.4\textwidth}
    \includegraphics[width=0.4\textwidth]{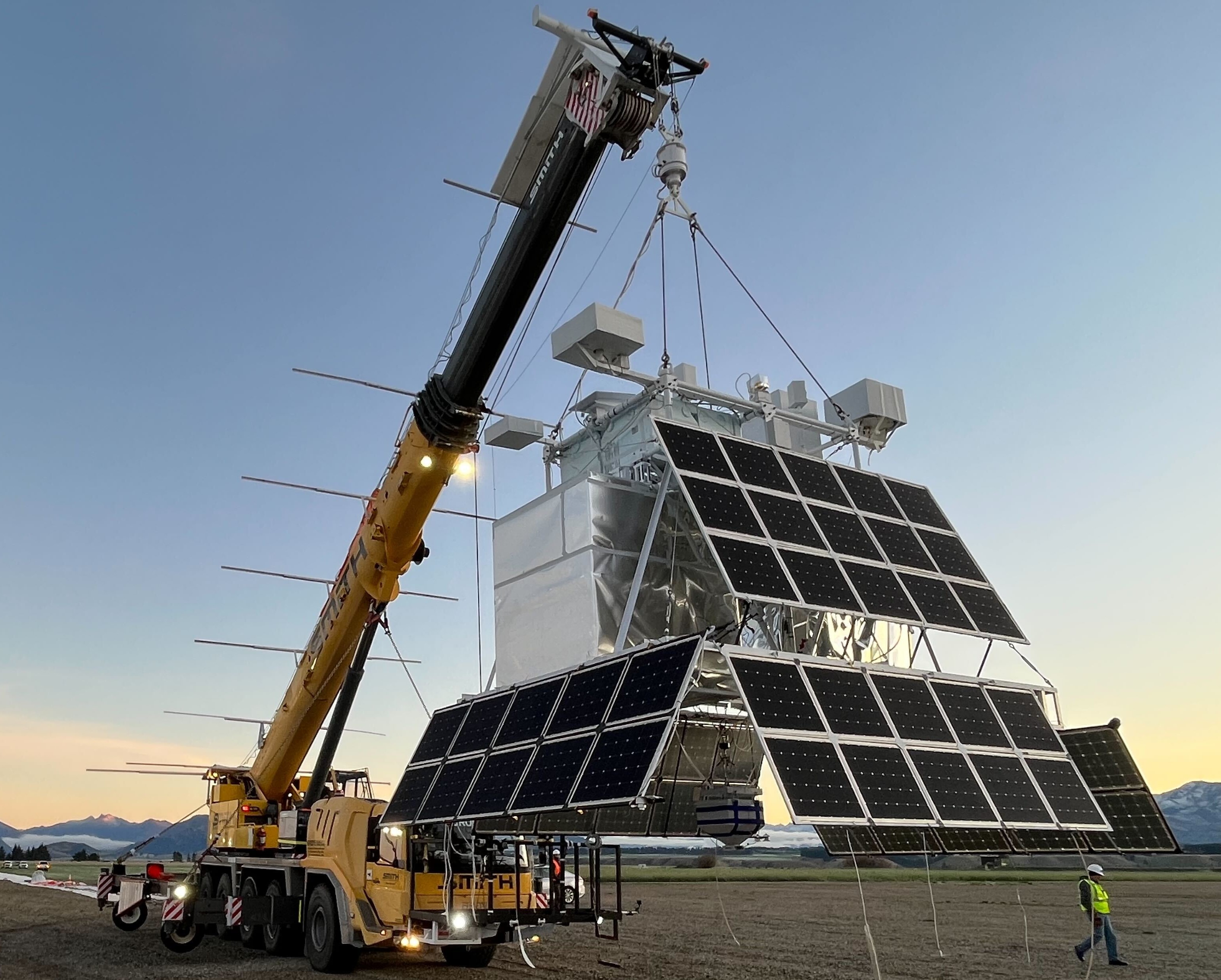}
    \caption{Entire payload hanging off the launch crane, attached to the balloon minutes before takeoff.}
    \label{fig:gondola_launch}
\end{wrapfigure}

As a precursor for POEMMA and Terzina, we developed a small Cherenkov telescope (CT) as part of the Extreme Universe Space Observatory Super Pressure Balloon 2 (EUSO-SPB2). Figure 1, shows the EUSO-SPB2 payload right before launch in Wanaka, New Zealand. EUSO-SPB2 was successfully launched on May 13 but was terminated only 36 hours after the launch because of a leak in the balloon. The payload was lost after it sank into the South Pacific. Nevertheless, the EUSO-SPB2 CT  demonstrated the primary mission goal, of the IACT from a suborbital platform. Our CT telescope operated for two nights during which we recorded 10\, GB of data in addition to engineering and instrument slow-control data.

We recorded, for the first time, candidates of cosmic-ray induced air-showers at a float altitude of about 33\, km by tilting the telescope above the Earth's limb. We also collected a comprehensive data set to study background light intensities and background events. We recorded trigger rate curves and other instrument performance data. On the second night, following notification that the balloon was leaking, our focus was on collecting as much data as possible, observing above and below the limb before the flight was terminated. The types of data collected during the flight include: 
\vspace{-.35cm}
\begin{enumerate}[noitemsep]
    \item  HLED flashes, LED flashes at 5 intensities throughout the regular observation, to monitor the performance of the instrument.
    \item  Night Sky Background (NSB) observations at varying azimuthal and elevation pointing directions and altitudes.
    \item Bifocal observations, including cosmic ray EAS candidates.
\end{enumerate}
\vspace{-.35cm}

The CT observations are pioneering the IACT for the first time at suborbital altitudes. In what follows, we present the design of the CT and discuss the telescope performance.

\section{CT Design Overview}
\vspace{0.1cm}
\begin{figure}[!hb]
    \includegraphics[width=1\textwidth]{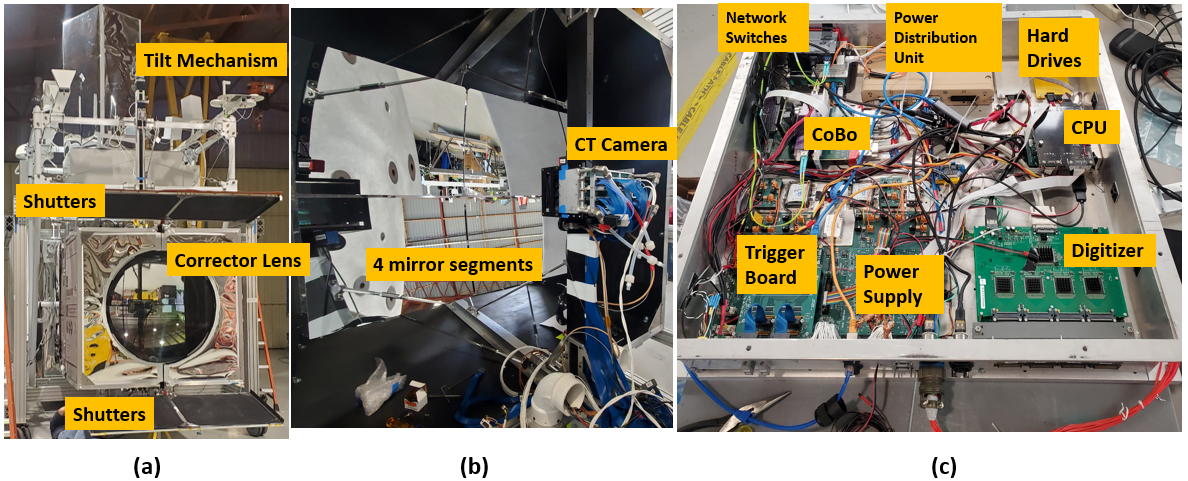}
    \caption{The CT during preparation for launch in Wanaka, New Zealand.}
    \label{fig:assembled_components}
\end{figure}
The Cherenkov Telescope (CT) is shown with open shutters in Figure \ref{fig:assembled_components}a. The Schmidt optics has about 1\,m diameter light collection area stemming from four mirror segments (see Figure \ref{fig:assembled_components}b). The four mirror segments are split into two pairs. Each pair focuses the light from a source at infinity at a spot separated by $10$\, mm or $0.8^\circ$ from the spot produced by the other mirror pair in the focal plane. This bifocal alignment of the mirrors effectively implements the so-called stereo trigger. It helps discriminate accidental triggers due to fluctuations in the night-sky background. For SPB2 it distinguishes between light coming from outside the telescope that is focused to make two spots and direct hits on the camera by low energy cosmic rays that can trigger just one pixel.

The CT camera is located on a curved focal surface, shown in Figure \ref{fig:assembled_components}b. The camera has 512 silicon-photomultiplier pixels and a $6.4^\circ \times 12.8^\circ$(vertical and horizontal) field of view. The camera pixels are grouped in $4\times 4$ arrays constituting 32 camera modules. For the SiPMs we used the S14521-6050AN-04 from Hamamatsu, a 6\,mm $\times$ 6\,mm SiPM that comes assembled in a $4\times4$ matrix, i.e. one camera module consists of one SiPM matrix. We chose the S14521 for its broad sensitivity from 200 nm to 1000 nm and its peak photon detection efficiency of $50\%$ at 450 nm.

Each SiPM matrix is connected to a custom-designed Sensor Interface and Amplifier Board (SIAB) populated with two Multipurpose Integrated Circuit (MUSIC) chips \cite{Gomez}. The MUSICs shape and amplify the SiPM signals, which are then guided to a custom backplane. From there, cables carry the signals to the digitizer boards inside the main electronics box (MEB), which is mounted below the telescope and shown in Figure \ref{fig:assembled_components}c. The digitizer is the ASIC for General Electronics for TPC (AGET) \cite{POLLACCO201881}. The AGET samples the signals with a rate of 100MSa/s and a buffer depth of 5.12$\mu s$. Upon receiving a readout command from the trigger board, the signals are digitized with 12-bit resolution. The digitized signals are managed by a Concentration Board (CoBo). The CoBo applies time stamps, performs zero suppression, and compresses the digitized signal. The CT camera and readout are commanded through the CT-CPU located inside the MEB, which also stores the data before it is downloaded. Data was downloaded during the flight using Starlink antenna and satellite system. The CT design is detailed in \cite{ICRC2021_CT_Overview}.

The telescope is suspended in the gondola of EUSO-SPB2 on a pivot arm at its back and can be tilted up and down between $3.5^\circ$ and $-13^\circ$ with respect to the horizontal at a rate of $\sim 0.8^\circ/min$ by a linear stage, attached near the front of the telescope. The gondola is attached to the rest of the balloon flight train through a lightweight rotator, which allows the rotation of the entire payload in azimuth.

\section{In-Flight Performance of the CT Instrument}
\begin{figure}
    \centering
    \includegraphics[width=1\textwidth]{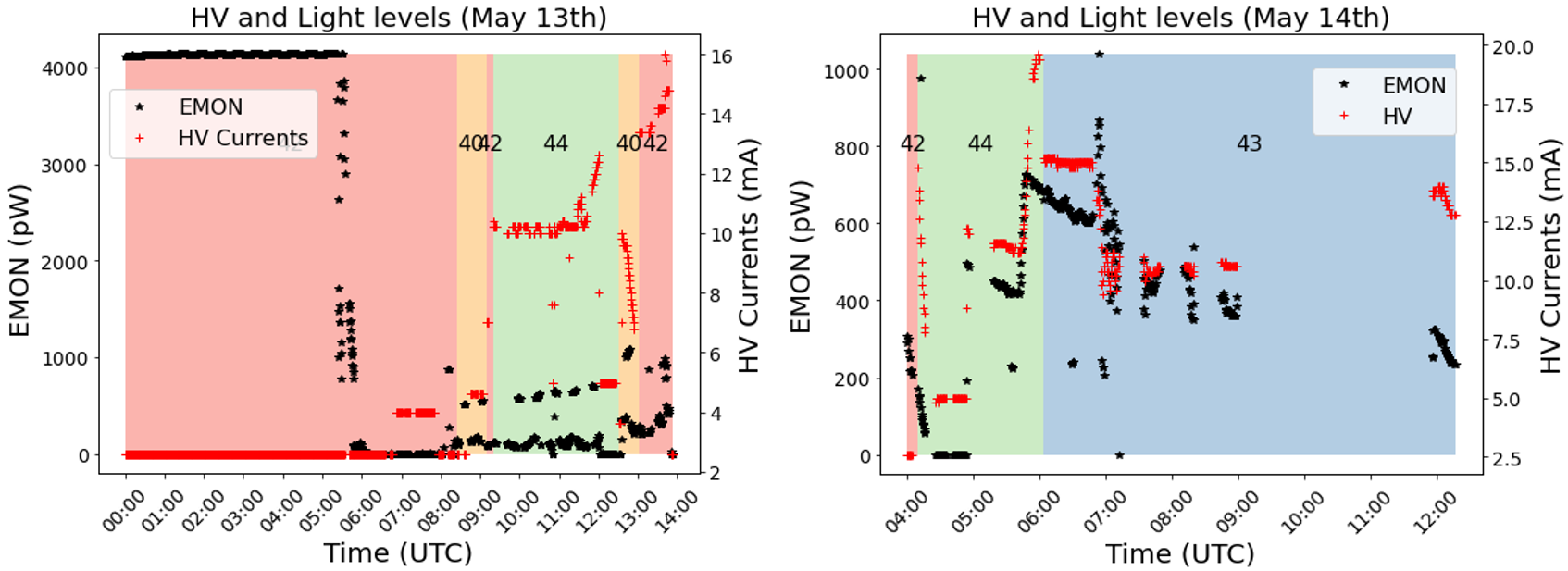}
    \caption{The plots show the summary of SiPM bias voltage settings (colored rectangles), SiPM current (y-axis on the right), and background light levels (y-axis on the left) during the flight. At the beginning of the flight, light levels are at maximum, indicating daylight, then they continued to drop and vary depending on pointing direction and moonlight levels. Current levels roughly follow the amount of light but are also dependent on biased voltage settings. The biased voltage was adjusted based on brightness/current levels.}
    \label{fig:HV}
\end{figure}

\subsection{System Level Performance}
During the flight, we monitored temperatures of critical electronic components in the camera and the MEB as well as the SiPMs. We also logged the different voltages and currents delivered to the system. Status messages were transmitted to the ground every 30 seconds whenever communication with the CT-CPU was established. All status messages indicated nominal instrument behavior throughout the flight. 
\begin{wrapfigure}{R}{0.55\textwidth}
    \centering
    \includegraphics[width=.54\textwidth]{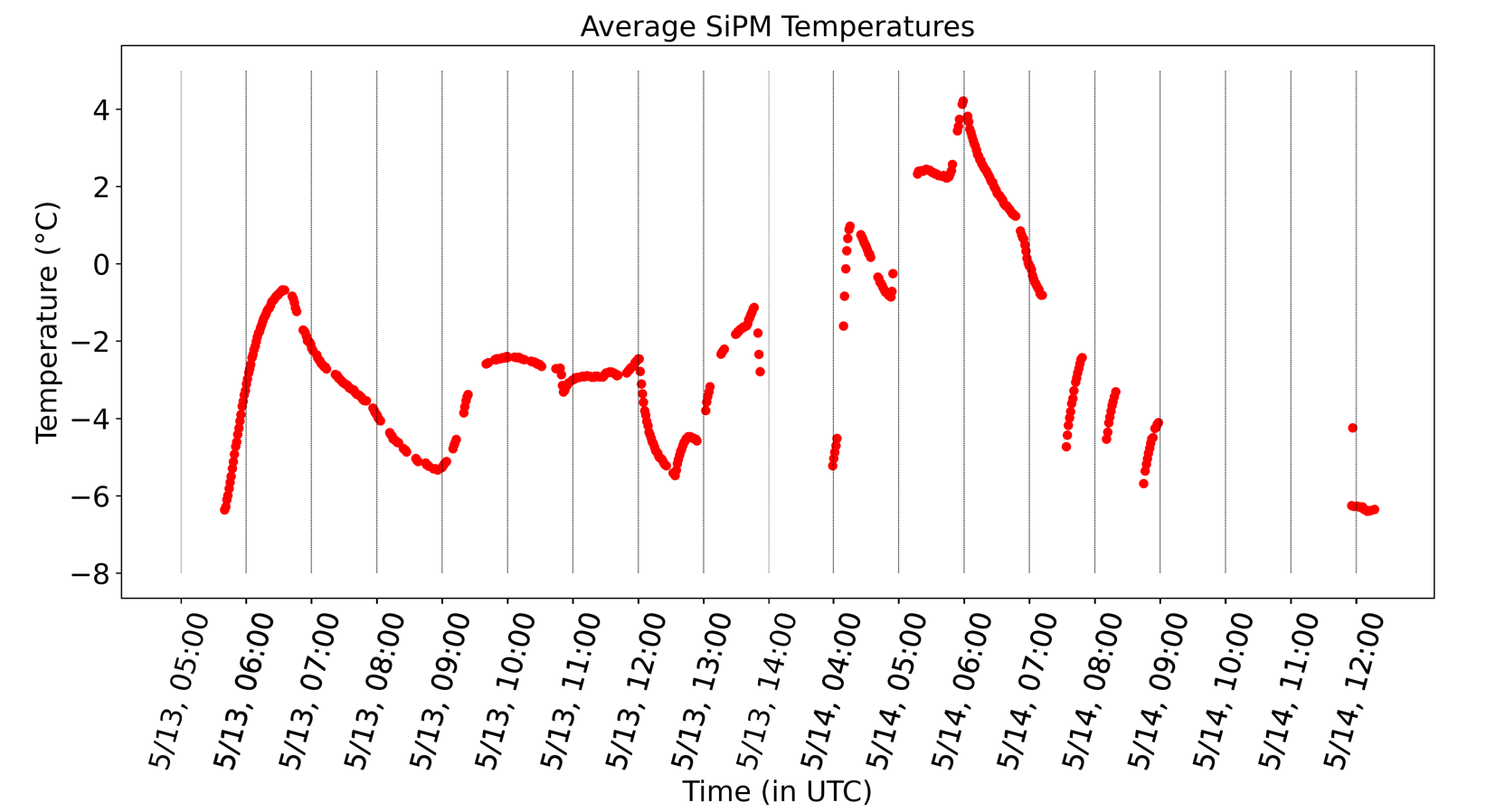}
    \caption{Average SiPM temperatures throughout the flight.}
    \label{fig:temp}
\end{wrapfigure}

The CT camera consumed 32W of power and was located inside a half inch foam "clam shell" enclosure. To regulate the resulting camera temperatures, an active cooling system was integrated, consisting of heat pipes thermally connecting the MUSIC chips with cold plates fixed to the sides of the camera. Two connected gear pumps circulated antifreeze from the heat-dissipating cold plates to radiators fixed to the exterior of the telescope housing, in equilibrium with space. During the flight, the pumps ran at a higher-than-anticipated speed, which we attribute to an unintended shift of the voltage controlling the pump speed. Still, the cooling system performed well and kept the camera electronics and SiPMs between $-6^\circ$C and $4^\circ$C during observations, minimizing the potential noise that heat would generate in the SiPM signals. Figure \ref{fig:temp} shows the SiPM temperatures averaged over the entire camera for the two nights of operation.

The readout electronics and the camera electronics behaved as expected with the exception of the CoBo board. During the descent of the balloon through the Tropopause and the upper Troposphere, the temperature of the electronics inside the MEB dropped below design values as convective cooling became significant. The CoBo board stalled, making it impossible to collect data. However, we resolved this by adding delays in the data acquisition software in real time through a Starlink connection. A Starlink antenna was used with the system to download science data from the telescope.

In addition to basic status monitoring, instrument data includes the bias voltage and current for each camera module, trigger rates during each data run, and background light levels. The bias voltage of the SiPMs was set between 40V and 44V. A lower bias was chosen when the ambient light increased due to the rising moon. We were able to operate the CT camera even with a 30\% moon phase, pointed away from the moon. A timeline of SiPM bias voltage, uncalibrated background light levels, and SiPM current is displayed in Figure \ref{fig:HV}.
\vspace{-4pt}
\subsection{HLED Measurements}
\vspace{-5pt}
\begin{figure}[!htbp]
    \centering
    \includegraphics[width=1\textwidth]{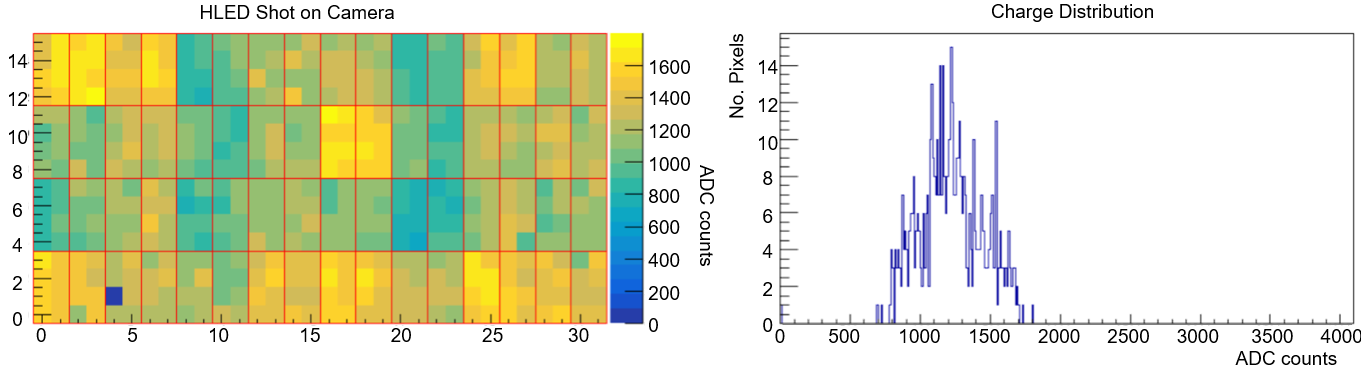}
    \setlength\abovecaptionskip{-0.7\baselineskip}
    \caption{HLED scan at highest amplitude. Left: Focal surface of the camera, each small square is a pixel and shows the light amplitude in ADC counts, each red rectangle is a MUSIC chip. One dark blue pixel was not operating during this measurement. Right: charge distribution across all pixels during this HLED scan.}
    \label{fig:HLED}
\end{figure}
\begin{wrapfigure}{R}{0.55\textwidth}
\vspace{-.45cm}
      \captionsetup{width=0.5\textwidth}
    \centering
    \includegraphics[width=.5\textwidth]{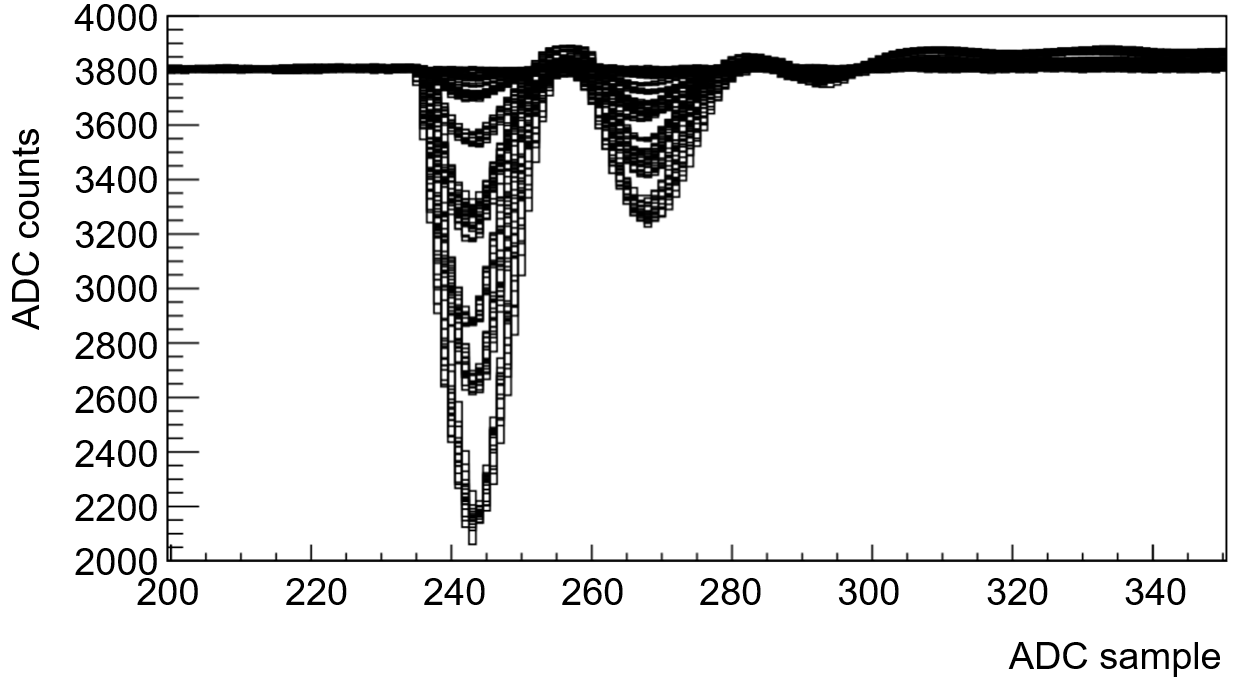}
    \caption{HLED traces averaged over all camera pixels for individual HLED flashes. Clearly identifiable are the five different HLED intensities. There are 208 HLED flashes recorded during the first night.}
    \label{fig:HLED_traces}
\end{wrapfigure}
Figure \ref{fig:HLED} shows the camera response to one health LED (HLED) flash. The HLEDs are located right above and below the camera pointing at the mirrors. The visible structure in the camera response is due to the gaps between the mirror segments. We recorded about 32,000 HLED flashes throughout the flight tracking the stability of the camera response. The HLED flashed at 5 different intensities. The HLED data will be used in the ongoing data analysis to calibrate and flatfield the camera signals.

Figure \ref{fig:HLED_traces} shows the overlayed traces of all 208 HLED flashes recorded in the first night averaged over all camera pixels. No calibration factors have been applied. Clearly identifiable are the five HLED intensities. The narrow bands qualitatively demonstrate camera stability in amplitude and time.

\section{Preliminary Results}

Besides camera performance data we also successfully collected, images of the Night Sky Background (NSB) at various telescope conditions and configurations such as biased voltage, tilt, and orientations in azimuth. Further, we collected bifocal events, that are likely to be candidates of cosmic rays-induced EAS, neutrino-induced EAS, direct hits, or background events that will need to be studied more in-depth. The below section describes some example events that we have identified in our data set by manually scanning events. 
\subsection{Night Sky Background}
The Night Sky Background (NSB) is the ambient background photon field which, during dark, moonless nights, is dominated by airglow, direct and scattered starlight, and zodiacal light \cite{BENN1998503}. This is the first time the NSB has been observed with a Cherenkov Telescope from an altitude of 33\,km pointing above and below the Earth's limb. We are in the process of analyzing the brightness of the sky at various tilt angles and characterizing any events the CT detected.
\subsection{Sample Bifocal Events from Recorded Air Showers}
Another data type collected was during instances when the telescope was directed toward or below the limb in order to search for PeV-energy Earth-skimming neutrinos. We focused, particularly, on tau-neutrinos penetrating the Earth at shallow angles, interacting and producing tau-leptons emerging from the ground. As taus decay within the atmosphere, they initiate extensive air showers (EAS) characterized by billions of charged particles. Subsequently, a small portion of the particles' energy is converted into optical Cherenkov emission radiated within a narrow cone centered around the shower axis \cite{AlvarezMuniz2019, Cummings2021}. By pointing the Cherenkov telescope below the horizon we optimize the acceptance for tau-neutrino-induced air showers. Offline analysis is in progress to search for all possible triggered instances of such showers.

When the telescope was pointed above the Earth's limb, the Cherenkov telescope was able to record images of cosmic-ray-induced Air-Shower images in the upper atmosphere. And indeed, by scanning through the data we can easily identify events, which have all the characteristics of air shower images. The recorded images show similar optical characteristics to events recorded during field tests when the telescope was pointed upwards to search for TeV scale EASs. An in-depth analysis employing proper event classification algorithms is in progress.
\begin{figure}
    \centering
    \includegraphics[width=1\textwidth]{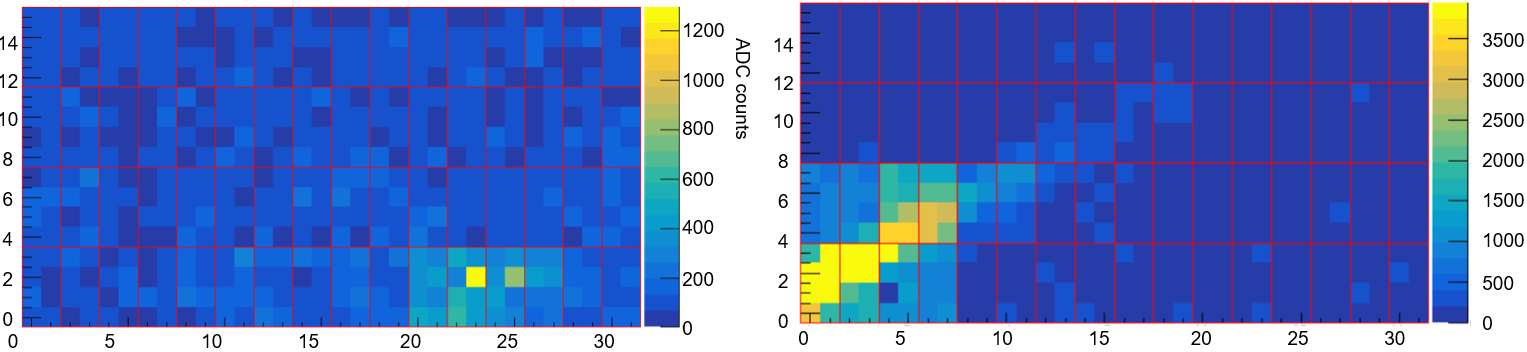}
    \caption{Candidates for Cosmic ray induced air shower events, Left: BiFocal, Right: Cherenkov cone. Both images were taken as the focal plane was pointed at the limb, therefore the top half of the display is below the Earth's limb while the bottom half is above, the image is flipped due to optics.}
    \label{fig:bifocal_cr}
\end{figure}
In total, we triggered on 31,269 events. The majority of these are accidental triggers, about 45min of observation was pointing above the limb for cosmic ray EAS detections, and some events were recorded while pointing below the limb but the top row of the camera still pointing at or above.  Figure \ref{fig:bifocal_cr} shows an example of two bifocal events detected on the second night of observations while pointing the telescope above the Earth's limb. The event in the left picture is an example bifocal event, while the event on the right has an extended topology over several degrees in length. Both events are likely the result of cosmic ray-induced EASs.

\subsection{Sample “other” results}
\begin{figure}
    \centering
    \includegraphics[width=1\textwidth]{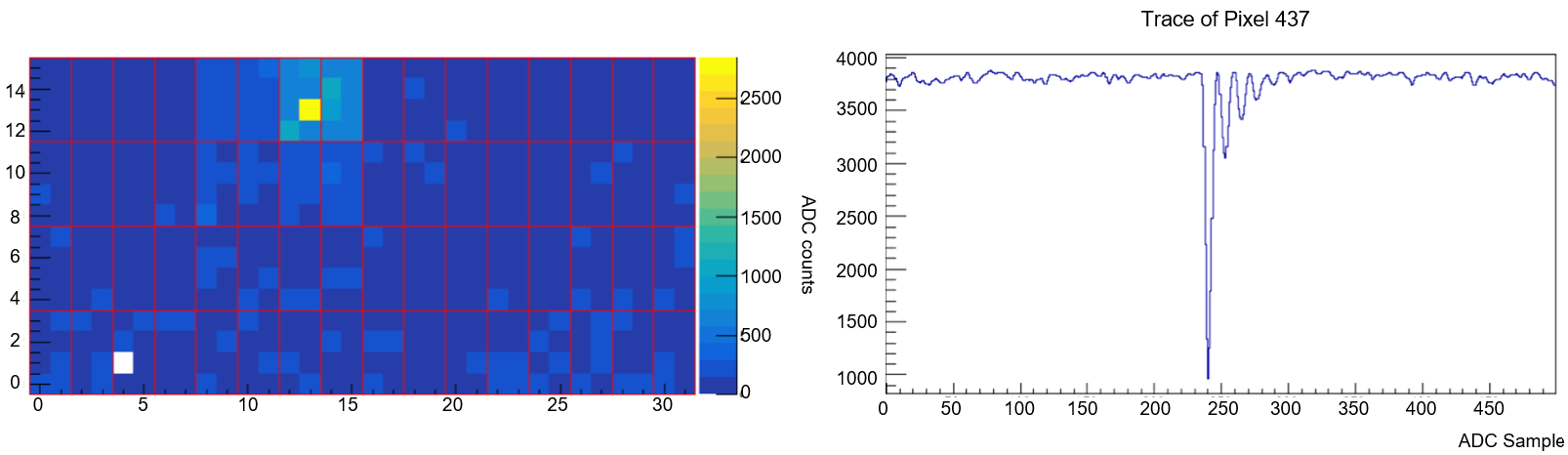}
    \caption{Possible direct cosmic-ray hit}
    \label{fig:wierd_events}
\end{figure}

Based on initial scans through detected events, we have also noticed events that do not show the expected bifocal topology. Figure \ref{fig:wierd_events} shows one example where one pixel exhibits an intense peak of about 3000 ADC counts which can be converted to about 300 photo-electrons and another pixel in the neighboring MUSIC is also above the trigger threshold within the 50\,ns coincidence window. The event is likely due to a direct cosmic ray hit. Due to the large signal amplitude, it is unlikely to be caused by an NSB fluctuation.
\vspace{-2mm}
\section{Conclusion}
For the first time, we operated a Cherenkov Telescope from suborbital altitudes. The balloon launched on May 13th and flew for 36 hours, taking data in the dark for about 14 hours. Despite the short flight, we collected a sufficient amount of data to study the performance of the camera. During the two nights, we verified the functionality of the CT camera, recorded trigger rates, took HLED data, and took data by pointing at various positions in the sky with and without moonlight present. While a detailed in-depth analysis of the data is still pending, we can already verify that the camera performed to expectation and air-shower images have been recorded when pointing the telescope above the Earth's limb, including candidates for cosmic-ray induced EAS. In conclusion, the CT performed very well and can be implemented in future  experiments such as POEMMA and Terzina, which will give us a new view into the UHE neutrino and VHE cosmic ray window.
\section{Acknowledgements} 
\noindent
The authors acknowledge the support by NASA awards 11-APRA-0058, 16-APROBES16-0023, 17-APRA17-0066, NNX17AJ82G, NNX13AH54G, 80NSSC18K0246, 80NSSC18K0473,   \\
80NSSC19K0626, 80NSSC18K0464, 80NSSC22K1488, 80NSSC19K0627 and 80NSSC22K0426, the French space agency CNES, National Science Centre in Poland grant n. 2017/27/B/ST9/02162, and by ASI-INFN agreement n. 2021-8-HH.0 and its amendments. This research used resources of the US National Energy Research Scientific Computing Center (NERSC), the DOE Science User Facility operated under Contract No. DE-AC02-05CH11231. We acknowledge the NASA BPO and CSBF staffs for their extensive support. We also acknowledge the invaluable contributions of the administrative and technical staffs at our home institutions.
\bibliographystyle{JHEP}
\bibliography{references}

\newpage
{\Large\bf Full Authors list: The JEM-EUSO Collaboration\\}

\begin{sloppypar}
{\small \noindent
S.~Abe$^{ff}$, 
J.H.~Adams Jr.$^{ld}$, 
D.~Allard$^{cb}$,
P.~Alldredge$^{ld}$,
R.~Aloisio$^{ep}$,
L.~Anchordoqui$^{le}$,
A.~Anzalone$^{ed,eh}$, 
E.~Arnone$^{ek,el}$,
M.~Bagheri$^{lh}$,
B.~Baret$^{cb}$
% ...
}
\end{sloppypar}
\vspace*{.3cm}

{ \footnotesize
\noindent
% ...
$^{cb}$ Universit\'e de Paris, CNRS, AstroParticule et Cosmologie, F-75013 Paris, France\\
% ...
$^{ed}$ Istituto Nazionale di Fisica Nucleare - Sezione di Catania, Italy\\
% ...
$^{eh}$ INAF - Istituto di Astrofisica Spaziale e Fisica Cosmica di Palermo, Italy\\
% ...
$^{ek}$ Istituto Nazionale di Fisica Nucleare - Sezione di Torino, Italy\\
$^{el}$ Dipartimento di Fisica, Universit\`a di Torino, Italy\\
% ...
$^{ep}$ Gran Sasso Science Institute, L'Aquila, Italy\\
% ...
$^{ff}$ Nihon University Chiyoda, Tokyo, Japan\\ 
% ...
$^{ld}$ University of Alabama in Huntsville, Huntsville, AL, USA\\
$^{le}$ Lehman College, City University of New York (CUNY), NY, USA\\
% ...
$^{lh}$ Georgia Institute of Technology, USA\\
% ...
}

\end{document}